# On the role of intrinsic disorder in the structural phase transition of magnetoelectric EuTiO$_3$

Mattia Allieta[1], Marco Scavini[1,*], Leszek Spalek[2,3], Valerio Scagnoli[4,5], Helen C. Walker[5], Christos Panagopoulos[2,3,6], Siddharth Saxena[2], Takuro Katsufuji[7] and Claudio Mazzoli[8,5].

[1] Dipartimento di Chimica Fisica ed Elettrochimica, Università degli Studi di Milano, Via Golgi 19, 20133 Milano, Italy;

[2] Cavendish Laboratory Madingley Road Cambridge CB3 0HE United Kingdom;

[3] Department of Physics, University of Crete and FORTH, GR-71003 Heraklion, Greece;

[4] Swiss Light Source, Paul Scherrer Institut, CH-5232 Villigen PSI, Switzerland;

[5] European Synchrotron Radiation Facility, 6 rue Jules Horowitz, BP 220, 38043 Grenoble Cedex 9, France;

[6] Division of Physics and Applied Physics, Nanyang Technological University, Singapore;

[7] Department of Physics, Waseda University, Tokyo 169-8555, Japan.

[8] Politecnico di Milano, p.zza L. Da Vinci 32, I-20133, Milano, Italy.

[*] Corresponding author, marco.scavini@unimi.it

**Abstract**

Up to now the crystallographic structure of the magnetoelectric perovskite EuTiO$_3$ was considered to remain cubic down to low temperature. Here we present high resolution synchrotron X-ray powder diffraction data showing the existence of a structural phase transition, from cubic *Pm*-3*m* to tetragonal *I*4/*mcm*, involving TiO$_6$ octahedra tilting, in analogy to the case of SrTiO$_3$. The temperature evolution of the tilting angle indicates a second-order phase transition with an estimated $T_c$=235K. This critical temperature is well below the recent anomaly reported by specific heat measurement at $T_A$~282K. By performing atomic pair distribution function analysis on diffraction data we provide evidence of a mismatch between the local (short-range) and the average crystallographic structures in this material. Below the estimated $T_c$, the average model symmetry is fully compatible with the local environment distortion but the former is characterized by a reduced value of the tilting angle compared to the latter. At $T$=240K data show the presence of local octahedra tilting identical to the low temperature one, while the average crystallographic structure remains cubic. On this basis, we propose intrinsic lattice disorder to be of fundamental importance in the understanding of EuTiO$_3$ properties.

**Introduction**

Multiferroic materials attract a great deal of interest due to the complex phenomena arising from multiple coupled order parameters existing in a single system [1]. In the case of simultaneous ordering interplay, as in the subset of materials called magnetoelectrics (MEs), the control of ferroelectric polarization via a magnetic field [2] and of magnetic phases by an electric field [3], has been proved possible.

The interplay of spin and other electronic or lattice degrees of freedom can induce giant magnetoelectric effects [4-5], dynamic behavior [6], as well as novel types of excitations [7], paving the way for future applications in sensors, data storage and spintronics [8-9]. In this paper we present the case of magnetoelectric $EuTiO_3$ (ETO) showing an unusual interplay between dielectric, magnetic and structural degrees of freedom. At room temperature (RT) its crystal structure has been reported to be *Pm-3m* and no phase changes have been observed to occur down to 108K [10], as deduced from lab source powder X-ray diffraction. From the dielectric point of view, ETO is described as a quantum paraelectric, as its low temperature dielectric constant increases on cooling and saturates below approximately 30K [11]. No long range polarization is known to set in, despite high values of susceptibility, typical of a paraelectric state stabilized by quantum fluctuations [12]. The localized 4f moments on the $Eu^{2+}$ sites order at $T_N$ = 5.3K in a antiferromagnetic arrangement [13]. Concomitant with the onset of antiferromagnetism the dielectric constant decreases abruptly (by $\varepsilon'_{5.5K}/\varepsilon'_{2K}$~3.5%) and shows a strong enhancement as a function of the applied magnetic field (~7% at B~1.5T), providing evidence for the magnetoelectric coupling [11]. In bulk MEs the coupling between various degrees of freedom is realized at a microscopic level [3], hence the crystallographic structure of ETO as a function of temperature is vital to any further investigation and modeling. In this paper, we report on the structure of ETO at low temperature, as given by high resolution synchrotron X-ray powder diffraction analysis. Since the diffraction experiments were performed, the authors have been made aware of a recent publication, reporting about a specific heat anomaly detected in ETO powders at high temperature [14], and in our discussion we address the differences arising from different experimental probes.

**Experiments**

High quality ETO samples have been grown by using the floating-zone method as outlined in [15]. The growth procedure involves melting a pressed rod of mixed starting materials ($Eu_2O_3$, Ti and $TiO_2$) under an Ar atmosphere inside a floating-zone furnace. Polycrystalline samples coming from the same batches as the one used for X-ray measurements were checked by specific heat first, showing an anomaly identical to the one recently reported in literature [14]. ETO crystals extracted from the inner part of the grown crystalline rod were powdered, loaded in a 0.70 mm diameter capillary and spun during measurements to improve powder randomization. A wavelength of $\lambda=$ 0.34986(1) Å was selected by using a double-crystal Si(111) monochromator. Several samples were checked and a few selected on the basis of RT X-ray powder diffraction (XPD) measurements. The diffracted rays FWHM and symmetry criteria were applied to select a couple of best samples: in the following we report about measurements performed on those specimens. Different data collection strategies were employed: (i) in the $0 \leq 2\vartheta \leq 60°$ range data were collected for a total counting time of 2 hours at room temperature (RT), 240K, 230K, 215K, 200K, 175K, 160K, 140K, 120K, 100K, 80K; (ii) in the $3 \leq 2\vartheta \leq 15°$ range 30 XPD patterns were collected while sweeping the temperature from 300 K to 200 K; (iii) at 100 and 240 K data were collected by summing several scans for a total counting time of 7 hours ($Q_{max} \sim 27$ Å$^{-1}$.) to achieve the necessary quality for Pair distribution function analysis.

The temperature on the sample was varied using a $N_2$ gas blower (Oxford Cryosystems) mounted coaxially to the sample capillary, being orthogonal to the scattering plane.

Diffractograms were indexed by using the DICVOL91 software [16]. Le Bail-type and Rietveld refinements were performed using the GSAS program [17]. In particular in the Rietveld refinement the background was fitted by Chebyshev polynomials. The absorption correction was performed through the Lobanov empirical formula [18] implemented for the Debye-Scherrer geometry. In the last refinement cycles, scale factor, cell parameters, positional coordinates and isotropic thermal parameters were allowed to vary as well as background and line profile parameters.

**Rietveld analysis**

At room temperature ETO is isostructural to SrTiO$_3$ (STO, space group $Pm$-$3m$, $a$ = 3.905 Å [20]), and the Rietveld refinement of XPD patterns by the same cubic structural model [10, 19] leads to a satisfactory description of our data. Our lattice parameters agree well with the literature (this work: $a$ = 3.904782(5)Å; Ref.[19]: $a$ = 3.904Å). In Fig. 1 (*a*), selected portions of the XPD patterns collected at various temperatures are shown. The contrast between the unperturbed (111) reflection family and the (200) split one is evident. In particular for this last reflection family, the intensity ratio of the two split peaks is ~1/2, suggesting a cubic to tetragonal structural phase transition.

To solve the low temperature structure we have concentrated on XPD data collected at 100K. First of all, 20 strong independent peaks were indexed, resulting in a tetragonal unit cell of lattice parameters $a$, $b$ = 3.896 Å, $c$ = 3.903Å. Then, by using a Le Bail-type profile-matching without structural model, based on the holohedral space group $P4/mmm$, we detected the presence of a few non indexed weak reflections as shown in Fig. 1 (*b*). A search for possible supercells gave a unit cell with lattice parameters of $a$, $b$ = 5.509Å, c = 7.808Å and the analysis of systematic extinctions for this cell was compatible with a body centered lattice. The two previous tetragonal cells are related according to the transformation matrix: (1,1,0), (-1,1,0), (0,0,2) or, alternatively, the low temperature tetragonal supercell can be classified as a pseudocubic cell with unit cell metric $\sqrt{2}a \times \sqrt{2}a \times 2a$, where $a$ is the primitive cubic perovskite lattice parameter. By using the Le Bail-type refinement procedure we fitted the diffractograms acquired below room temperature with the new cell metric. In Fig. 2, we present the refined lattice parameters as a function of temperature for ETO together with X-ray diffraction data obtained for STO [21], where a reduced tetragonal cell metric was used for comparison purposes between the cubic and tetragonal phases in the two systems.

Figure 2 clearly shows similarities between ETO and STO. The latter was shown to exhibit a second order displacive phase transition below $T_c$=106K [22], so quite naturally we started from this hypothesis in the analysis of ETO structural phase transition. The possible mechanisms accounting for such modulation of the aristotype perovskite structure (*Pm-3m*) can be generated by Jahn-Teller distortion [23], tilting of corner-linked BO$_6$ octahedral units [24-25], polar distortions [25] or higher order mechanisms coupling several of these [26]. Given the electronic state of Ti in ETO (3d$^0$) [27], we concentrated on octahedral tilting and group-theoretical methods have been applied in order to classify compatible structures assuming a second order phase transition. The analysis yielded a list of 15 possible space groups [25] subgroups of the high temperature cubic one. Restraining the analysis to tetragonal space groups [25], the possible choices are (between parenthesis the related cell [28]): *I4/mmm* ($2a \times 2a \times 2a$), *P4/mbm* ($\sqrt{2}a \times \sqrt{2}a \times 2a$), *I4/mcm* ($\sqrt{2}a \times \sqrt{2}a \times 2a$) and *P4$_2$/nmc* ($2a \times a \times 2a$). Space groups related to a $2a \times 2a \times 2a$ cell, i.e. *I4/mmm* and *P4$_2$/nmc*, are incompatible with the data on the basis of the metrics (i.e. indexation of the peaks due to the supercell structure). Moreover *P4$_2$/nmc* has special extinction conditions, not fulfilled by the experimental pattern. The *P4/mbm* space group is completely ruled out by both the metrics and the lattice type. These arguments leave *I4/mcm* as the only possibility. The tetragonal structure in the *I4/mcm* space group is consistent with an out-of-phase tilting of octahedra around the tetragonal axis. The associated irreducible representation (irrep) is R$^+_4$ [25]. The direction of the distortion in the irrep space is indicated by the vector (*a*,0,0). To obtain the starting atomic positions in *I4/mcm* the ISOTROPY [29] package was used, on the basis of the Wyckoff sites occupied in the undistorted cubic structure: Ti on 1a (0, 0, 0), Eu on 1b (½, ½, ½) and O on 3d (½, 0, 0). The asymmetric unit of the *I4/mcm* subgroup consists then of Ti at 4c (0, 0, 0), Eu at 4b (0, ½, ¼), O1 at 4a (0, 0, ¼), O2 at 8h (*x*, *x* + ½, 0) with *x*~¼. Figure 3 shows the Rietveld refinement of data collected at 100K, as obtained by using the above structural model in the *I4/mcm* space group. In Table I, structural data and agreement factors obtained for patterns collected at different temperatures are listed. In the tetragonal phase we constrained both isotropic thermal parameters related to oxygen positions O1,

O2 to be the same. The thermal variation of the Ti-O and Eu-O bond lengths is given in Fig. 4 (*a*), (*b*). In the cubic phase the Ti-O1 distance decreases with decreasing temperature. Below 215K, according to the tilting of TiO$_6$ octahedra, Ti-O distances visibly split as a result of oxygen basal plane O2 displacement. Moreover it should be noted that since in the tetragonal phase the Ti-O1 indicates the distance between Ti and the apical O1 position of octahedron, its thermal variation follows the *c*-axis length evolution on cooling (see Fig. 2). The Eu-O1 distance remains approximately constant within the temperature range studied. In the tetragonal phase short (Eu-O2) and long (Eu-O2') distances are generated by the displacement of the O2 position. In perovskites, the out-phase tilting angle of TiO$_6$ octahedra ($\phi_-$), as calculated from the refined values of the x[O(2)] position, according to $\tan\phi_- = 1-4x[O(2)]$ [30], has been proposed as the primary order parameter of the *Pm-3m* to *I4/mcm* displacive transition of the average structural model. Figure 5 shows the values obtained from our refinements as a function of temperature, as listed in Table I. The temperature dependence of $\phi_-$ is expected to be described by a critical equation of the form: $\phi_-(T)=\phi_-(0)(1-T/T_c)^\beta$, with a critical exponent of $\beta = 0.5$ for a second-order phase transition. The correct determination of $T_c$ is fundamental for an effective estimation of the other model parameters. By setting $T_c$=215K on the basis of XPD evidence, we obtained the fit shown in Fig. 5 by the solid line, with parameters $\phi_-(0) = 4.03(2)°$ and $\beta = 0.40(1)$. The $\beta$ value thus obtained is different from the expected value. This finding is particularly important for internal consistency on the adopted procedure, based on our hypothesis of a second order phase transition.

Recently, an anomaly in the temperature dependence of specific heat measured on a powdered ETO sample has been reported, suggesting a structural instability at $T_A$=282(1)K [14]. Moreover, theoretical calculation performed by the same authors predicts a second-order antiferrodistortive phase transition which agrees perfectly with the *Pm-3m* to *I4/mcm* transition reported here. Despite this agreement, the discrepancy between the $T_c$ estimated from data in Fig. 5 (*a*) and the reported $T_A$=282(1)K [14] requires further investigation. For this reason we performed an accurate profile analysis of the (200) reflection indexed within a cubic unit cell (Fig. 1 (*a*)) on the 30 XPD patterns collected between 200 and 300K. In Fig. 6, we report the full width at half maximum (FWHM) of the (200) cubic reflection family as a function of temperature. In each pattern a single profile function was used because any attempt to describe the (200) peak by using a multiple peaks resulted in unphysical fluctuations of the fitting parameters. For sake of comparison, the temperature dependence of FWHM related to the (111) cubic reflection family is also shown. The FWHM of (200) smoothly increases on cooling below $T^*\sim235$K (Fig. 6). This suggests that the structural distortion inducing the FWHM variation occurs at higher temperature than the temperature estimated by the previous method. However, we point out that from the point of view of XPD at 235K the structural phase transition is just *incipient* without causing a detectable symmetry breaking until ~200K is reached. Indeed the splitting of cubic (200) peak is not visible in the 200K$\leq T \leq$235K range, the difference in the average (see below) *d*-space induced by the tetragonal distortion falling below the instrument resolution. It should be noted that a similar behavior has been already reported by some of the present authors concerning the tetragonal-to-orthorhombic transition of rare-earth cobaltite perovskite [31]. By setting $T_c$=235K, the fitting of the experimental $\phi_-(T)$ data by the same critical equation as before gives $\phi_-(0)$= 4.22(11)° and $\beta$= 0.53(7) as parameters best estimate. The resulting curve is shown in Fig. 5 by the dotted line. The value of the critical exponent is now close to the expected value, suggesting that our procedure is reasonably consistent. However $T^*\sim235$K is still too low to be consistent with the reported $T_A$ from specific heat measurements [14]. Such a discrepancy can arise from an incomplete description of the structural phase transition mechanism or because of intrinsic limits characteristic of the investigation technique. Among functional materials, like ETO [32], a general consensus is growing on the relation between physical properties of interest and disorder occurring at the local scale [33]. In the case of phase inhomogeneity, for example, the local and the average crystallographic structures are expected to differ, the correlation length (CL) of the structural distortion being spatially limited. If so, conventional analysis of XPD data, like Rietveld method, can be inadequate [34], being able to detect just long enough structural correlations (average structure). On the contrary, total scattering methods, like the pair distribution function (PDF), have been successfully applied to similar problems in Ti-based perovskites [35-36].

**PDF analysis**

We carried out a PDF analysis of the XPD data collected at $T$=100K, 240K. The PDF function $G(r)$ is obtained through the total structure factor $S(Q)$ via the sine Fourier Transform (FT):

$$G(r) = 4\pi r[\rho(r) - \rho_0] = \frac{2}{\pi} \int_{Q=0}^{Q_{max}} Q[S(Q) - 1]\sin(Qr)dQ, \quad (1)$$

where $Q_{max}=4\pi \sin\theta/\lambda$, $\rho(r)$ and $\rho_0$ are the local and average atomic number densities, $r$ is the interatomic distance and the total structure factor $S(Q)$ is obtained from the experimental coherent X-ray scattering intensity $I^{coh}(Q)$ according to:

$$S(Q) - 1 = \frac{I^{coh.}(Q) - <f^2(Q)>}{<f(Q)>^2}. \quad (2)$$

Here $f(Q)$ is the atomic scattering factor and the brackets < > stand for the average over the compound unit. To evaluate $I^{coh}(Q)$ consistently, the raw diffracted intensity $I(Q)$ collected at each temperature was corrected for background scattering, attenuation in the sample, multiple and Compton scattering. In particular, at high $Q$ the Compton scattering was removed by calculating the Compton profile with an analytical formula. In the middle-low $Q$ region the Compton scattering correction was applied by multiplying the calculated Compton profile with a monochromator cut-off function [34]. The corrected $I(Q)$ were then properly normalized, converted to get $S(Q)$ and Fourier transformed according to Eq. 1 to obtain the PDF data. The reduction operations have been done using the PDFGetX2 software [37]. Full structure profile refinements were carried out on PDF data using PDFgui program [38]. The program assesses the degree of accuracy of the refinement by the following agreement factor:

$$R_W = \left[ \frac{\sum w_i (G_i^{exp} - G_i^{calc})^2}{\sum w_i (G_i^{exp})^2} \right]^{1/2}. \quad (3)$$

Data collected were analyzed starting from $r=2.3$Å, i.e. excluding the shortest Ti-O distances. Indeed, the total $G(r)$ can be expressed as sum of partial $g_{i-j}(r)$ weighted for the atomic fractions and $f(Q)$ of the $i$ and $j$ components. Given the contrast between the X ray scattering factors of the element pairs involved, i.e. Eu-Eu, Eu-Ti, Eu-O, Ti-Ti, Ti-O, the partial $g_{Ti-O}(r)$ has the lowest weighting. Then, at very low $r$, the $G(r)$ peak related to Ti-O pair corresponds to very weak feature with respect to the baseline ($-4\pi r\rho_0$), as shown by the arrow in the inset of Fig. 7. PDF analysis is sensitive to different crystallographic CL via the refined range of the interatomic distance $r$. Figure 7 shows the full PDF profile structural refinement obtained at 100K by using the average $I4/mcm$ model in 2.3Å$\leq r \leq$20Å range. The agreement factor obtained (Rw=0.073) confirms the good quality of the fit at low enough temperatures when describing local distortions by the average model. Table II lists the structural parameter resulting from the PDF refinement at 100K. By considering the U(O1), U(O2) unrelated we obtain a marked improvement of the fit quality, so that all the reported PDF refinements were performed without oxygen thermal motion constraints. In the following the proposed structural order parameter $\phi_-$, as obtained by both Rietveld and PDF refinement, is compared as a function of $r$. As reported in Table II, for 2.3Å$\leq r \leq$20Å (short-range) PDF analysis gives a structural order parameter of $\phi_-=8.05(4)°$, while for 20Å$\leq r \leq$50Å (long-range) a value of $\phi_-=3.3(1)°$ is obtained, showing a strong dependence of the tilting angle as a function of the interatomic distance. In addition, the $\phi_-$ value found above 20Å is in close agreement with value obtained from Rietveld analysis of the XPD data ($\phi_-=3.14(8)°$). Figure 8 shows short-range portion of the PDF refinement at 240K. In panel *(a)* the average *Pm-3m* cubic model is shown. The calculated PDF systematically underestimates the intensity of most of the experimental peaks, proving the undistorted model to be inadequate. In panel (*b*), the fit performed by using the *I*4/*mcm* model in the same *r* range is shown. The *I*4/*mcm* has a better agreement with the data and the marked features in panel (*a*) are now well described. On the contrary, the *Pm-3m* structural model gives a reasonable fit over the long-range part (Table II) indicating that ETO completely recovers its average structure already at interatomic distances of ~20Å. This provides a clear evidence of a mismatch between the short and the long-range structure at a temperature as high as *T*=240K.

**Discussion**

PDF analysis of powder diffraction patterns suggests ETO to be an intrinsically disordered system as a clear mismatch between the short- and long-range crystallographic structures is evident at 240K. At 100 K the long-range tetragonal model describes the short-range PDF well qualitatively, but an increased value of the tilting angle is necessary to properly fit the data. From these results we propose a picture to reconcile the apparent discrepancy in the temperature anomalies *T*\* and $T_A$ as detected by non-local and local techniques, respectively. According to specific heat measurement interpretation [14], a second order phase transition occurs in ETO at $T_A$~282K. On the basis of the low temperature structural refinement shown above, we attribute this anomaly to a *Pm-3m* to *I*4/*mcm* displacive structural transition. The outcome of our PDF analysis shows how the CL of the tetragonal *I*4/*mcm* phase remains confined at the nanoscopic scale (~20Å) for ~235K≤*T*≤282K. The tilting changes randomly from one nanoregion to the adjacent, quickly averaging out the metric variation on a longer scale, i.e. the crystallite size, and thus reducing the average structure to a cubic space group. For this reason the local distortion cannot be detected by conventional techniques as the Rietveld analysis of XPD data. By decreasing *T*, a divergence of the tetragonal tilting CL takes place, resulting in a disorder-order transition at *T*\*~235K. Close to *T*\*, the magnitude of the tetragonal distortion (tilted) corresponding to the long-range ordered phase is still small, so that the transition shows up just as a weak feature in the temperature dependence of the cubic (200) FWHM. By further cooling the tetragonal deviation from the cubic metric increases, until the experimental resolution is finally sufficient to resolve it below 215K. There, splitting of the Bragg peaks as well as the appearance of superlattice reflections are clearly observed in Rietveld refinements. It is worth noticing that even at $T$=100K, i.e. well below $T^*$, the local tilting angle is greater than the one obtained from the long-range PDF refinement. At T=100K and 240K the local tilting angles (short-range refinement) are the same (~8°), see Fig. 5 and Table II, whereas the tilting angle obtained at T=100K from the long-range PDF refinement agrees well with the result of the Rietveld refinement for the average structure (~3°). On the other hand, $Cp$ measurements do not show a sharp feature but a quite broad one over the temperature range shown in [14], leaving the possibility of a further evolution of the structural distortion CL possible. In principle, the space group $I4cm$ obtained from the coupling of $R^+_4$ and the $\Gamma^+_4$ polar irreducible representation [25] could be compatible with our experimental results on the basis of metrics and extinction conditions [39]. In this case a strong dynamical behavior of the dielectric constant would be expected and given the mismatch between the local- and the long-range orders found in ETO, the system could act as relaxor ferroelectrics [40]. In these latter systems the disorder is typically introduced extrinsically, through chemical doping, while in ETO the structure itself seems to be willing to organize at a nanoscopic scale (forming domains of the order of ~20 Å).

**Conclusions**

In conclusion, in this paper we show that ETO undergoes a cubic to tetragonal structural phase transition below room temperature, on the basis of XPD data analysis. The *I4/mcm* space group generated by an out-of-phase tilting of TiO$_6$ octahedra gives the best description of our powder diffraction data at low temperatures. Literature specific heat measurement shows an anomaly at $T_A$=282(1) K. Our PDF analysis of XPD data shows that at *T*=240 K the structure of ETO is already distorted and consistent with the presence of local tilting regions embedded in a long-range cubic phase. From Rietveld analysis of XPD data the temperature dependence of the average tilting angle and of the (200) cubic peak FWHM suggests a second order transition taking place at $T^*$~235K. From the comparison between the Rietveld and the PDF analysis of XPD data, we propose that the difference between $T^*$ and $T_A$ is due to the CL scale evolution of the structural phase transition. At $T_A$ the cubic to tetragonal phase transition occurs at the nanoscale and it is then followed by a disorder-order transition at $T^*$, where the CL of the distorted regions starts to diverge, at least form the point of view of a non-local technique. Moreover, at T=100K the average model is consistent with the outcome of the long-range PDF refinement, while the short-range one suggests that a bigger value of the tilting angle is locally realized. This provides evidence of disorder at the nanometric scale even below $T^*$, suggesting ETO to be an intrinsically disordered system in which the structural phase CL changes dramatically over a wide range of temperatures. In case of a possible further symmetry breaking of the *I4/mcm* space group by inversion symmetry loss, fact that cannot be excluded by the present investigation, this material would represent the first evidence of an intrinsic relaxor magnetoelectric: disorder modulated interactions are expected to deeply influence the low temperature electric properties. This being the case, we believe that this fact has to be taken in suitable consideration when describing the peculiar properties of ETO as a quantum paraelectric material.

**Acknowledgments**

The authors gratefully acknowledge the European Synchrotron Radiation Facility for provision of beam time and Dr. Adrian Hill for assistance in using the ID31 beamline. We acknowledge also the financial support by the European Union through MEXT-CT-2006-039047 and EURYI research grants. The work in Singapore was supported by The National Research Foundation.

[39] By considering [001] as the polar direction and the Eu position (i.e. the $\Gamma_4^+$ polar irreducible representation) as a reference , we refined the data using the *I4cm* model. By comparing the results from *I4/mcm* and *I4cm* models we found no improvement in the refinement statistics at any temperature. However, it is worth noticing that the discrimination between 4/mmm and 4mm point symmetries requires a careful examination of the intensity distribution statistics, a very difficult task to be performed on powder diffraction data due to peak overlapping. Hence, the occurrence of inversion symmetry breaking cannot be unequivocally excluded by the current analysis of our high resolution XPD data.

# TABLES

Table I. Refined structural data of ETO obtained from synchrotron X ray powder diffraction.

| | RT | 240K | 230K | 215K | 200K | 175K |
|---|---|---|---|---|---|---|
| $a$ (Å) | 3.904782(5) | 3.902847(3) | 3.902521(2) | 3.901813(3) | 5.516890(7) | 5.515173(8) |
| $c$ (Å) | -------------- | -------------- | -------------- | -------------- | 7.804140(33) | 7.804589(34) |
| $U(Eu)(Å^2)$ | 0.00748(3) | 0.00655(2) | 0.00623(2) | 0.00593(2) | 0.00507(2) | 0.00471(2) |
| $U(Ti)(Å^2)$ | 0.00454(6) | 0.00400(6) | 0.00389(6) | 0.00373(6) | 0.00335(5) | 0.00315(5) |
| $U(O)(Å^2)$ | 0.0106(2) | 0.0105(2) | 0.0101(2) | 0.0107(2) | 0.0092(2) | 0.0090(2) |
| $x[O(2)]$ | -------------- | -------------- | -------------- | -------------- | 0.2439(7) | 0.2410(6) |
| $\phi$ (°) | -------------- | -------------- | -------------- | -------------- | 1.41(17) | 2.06(14) |
| Rwp | 0.1023 | 0.1073 | 0.1069 | 0.1100 | 0.0864 | 0.0945 |
| $R(F^2)$ | 0.0452 | 0.0422 | 0.0400 | 0.0427 | 0.0251 | 0.0268 |
| $\chi^2$ | 4.494 | 5.886 | 5.722 | 5.117 | 4.011 | 3.622 |

| | 160K | 140K | 120K | 100K | 80K |
|---|---|---|---|---|---|
| $a$ (Å) | 5.513909(8) | 5.512217(6) | 5.510639(6) | 5.509309(6) | 5.507642(9) |
| $c$ (Å) | 7.804941(28) | 7.805394(21) | 7.805601(6) | 7.805572(20) | 7.805161(25) |
| $U(Eu)(Å^2)$ | 0.00445(2) | 0.00407(2) | 0.00366(2) | 0.00325(1) | 0.00252(2) |
| $U(Ti)(Å^2)$ | 0.00302(5) | 0.00284(5) | 0.00270(4) | 0.00251(4) | 0.00220(4) |
| $U(O)(Å^2)$ | 0.0086(2) | 0.0084(2) | 0.0079(2) | 0.0074(2) | 0.0073(2) |
| $x[O(2)]$ | 0.2399(5) | 0.2384(4) | 0.2373(4) | 0.2363(4) | 0.2353(4) |
| $\phi$ (°) | 2.30(12) | 2.66(10) | 2.92(9) | 3.14(8) | 3.35(9) |
| Rwp | 0.0944 | 0.0942 | 0.0934 | 0.0945 | 0.0941 |
| $R(F^2)$ | 0.0273 | 0.0271 | 0.0274 | 0.0268 | 0.0258 |
| $\chi^2$ | 3.809 | 4.010 | 4.208 | 4.578 | 4.847 |

Table II. Refined structural parameters of ETO obtained from PDF refinements at 100K and 240K.

| Temperature | 100K | | 240K | | |
|---|---|---|---|---|---|
| Space Group | I4/mcm | | Pm-3m | | I4/mcm |
| Fitting $r$ range | 2.3Å≤$r$≤20Å | 20Å≤$r$≤50Å | 2.3Å≤$r$≤20Å | 20Å≤$r$≤50Å | 2.3Å≤$r$≤20Å |
| $a$/Å | 5.5124(1) | 5.5088(2) | 3.9019(5) | 3.9022(1) | 5.5197(8) |
| $c$/Å | 7.7931(8) | 7.8030(5) | --------------- | --------------- | 7.8002(2) |
| $U$(Eu)/Å$^2$ | 0.003225(6) | 0.003905(2) | 0.005463(2) | 0.006785(2) | 0.005794(2) |
| $U$(Ti)/Å$^2$ | 0.004317(6) | 0.005831(4) | 0.006056(5) | 0.007231(3) | 0.006228(5) |
| $U$(O1)/Å$^2$ | 0.00341(2) | 0.006988(3) | 0.02470(5) | 0.02594(2) | 0.00366(1) |
| $U$(O2)/Å$^2$ | 0.0429(4) | 0.0514(4) | --------------- | --------------- | 0.0449(2) |
| x[O(2)] | 0.2147(2) | 0.2356(6) | --------------- | --------------- | 0.2147(1) |
| $\phi$ (°) | 8.05(4) | 3.3(1) | --------------- | --------------- | 8.04(2) |
| Rw | 0.073 | 0.074 | 0.107 | 0.079 | 0.074 |

**FIGURES**

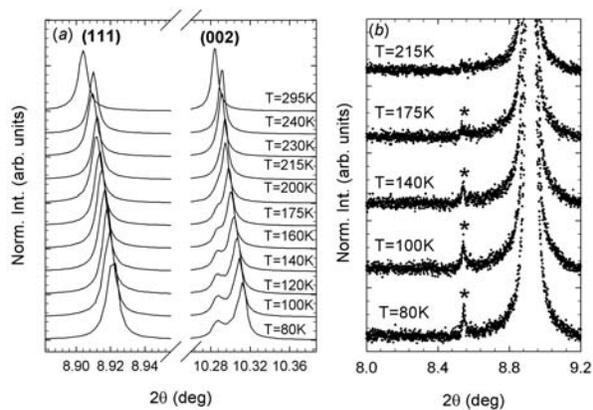

FIG. 1. Selected 2θ regions of ETO X-ray diffraction patterns are shown as a function of temperature. (*a*) shows the temperature evolution of the Miller index (111) and (002) peaks related to the RT cubic phase. (*b*) shows the appearance of a weak superlattice reflection as a function of temperature.

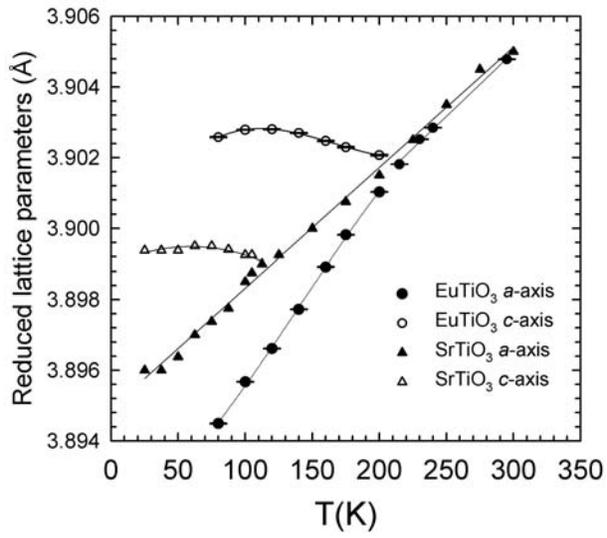

FIG. 2. Reduced lattice parameters of ETO and STO as a function of temperature. The full and opened circles are the *a* and *c*-axis of ETO. Full and opened triangles indicate the *a* and *c*-axis values of STO derived from Ref. [21]. The continuous lines are guides to the eye.

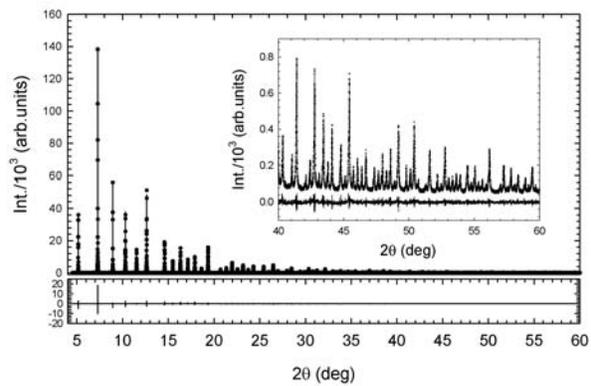

FIG. 3. Measured (dots) and calculated (lines) powder diffraction patterns for ETO at 100K. The inset shows a magnified view of the high angle diffraction peaks. The difference between the observed and fitted patterns is displayed at the bottom of each figure.

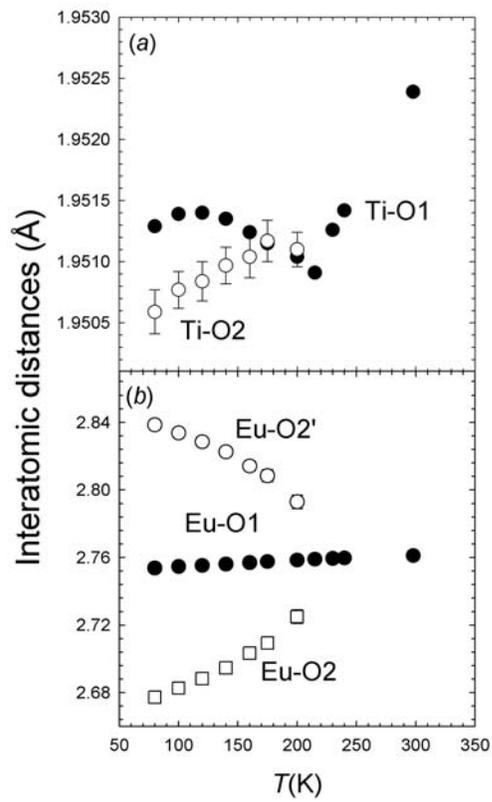

FIG. 4. (a), (b) show the refined Ti-O and Eu-O interatomic distances as a function of temperature.

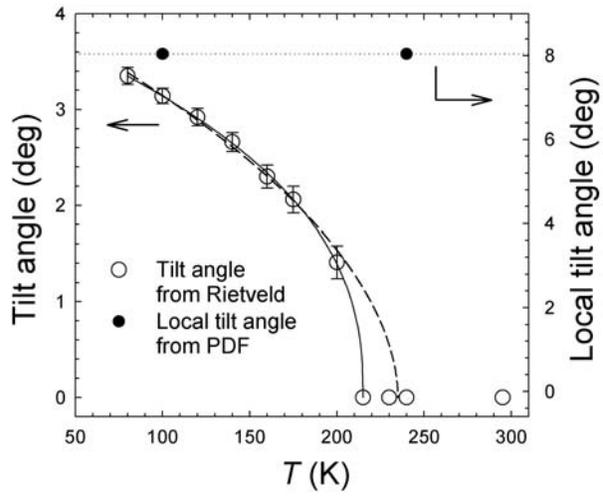

FIG. 5. ETO octahedral tilting angle values (empty circles) deduced from Reitveld analysis as a function of temperature (Table I). The continuous line is the best fit by the critical equation $\phi_{\_}(T)=\phi_{\_}(0)(1-T/T_c)^{\beta}$ by setting $T_c$=215K. The dotted line is the same but with $T_c$=235K. The full circles are the tilting angle values obtained from PDF refinements (Table II).

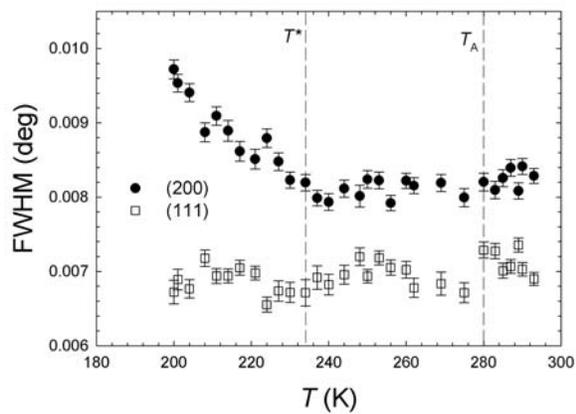

FIG. 6. Temperature evolution of the (200) and (111) Bragg families FWHM for the cubic phase of ETO.

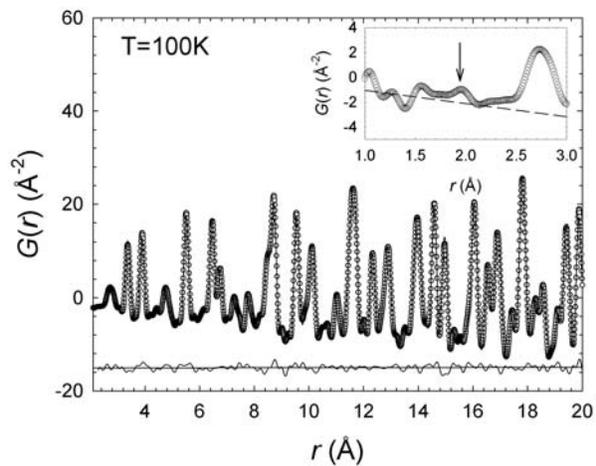

FIG. 7. Observed (dots) and calculated (continuous line) PDF obtained for ETO at 100K. The residual plot is shown at the bottom of the figure. In the inset a region of short interatomic distances is displayed. The dashed line is the baseline and corresponds to $-4\pi r\rho_0$ (see Equ.1). The arrow indicates approximately the $r$ position of Ti-O interatomic distances.

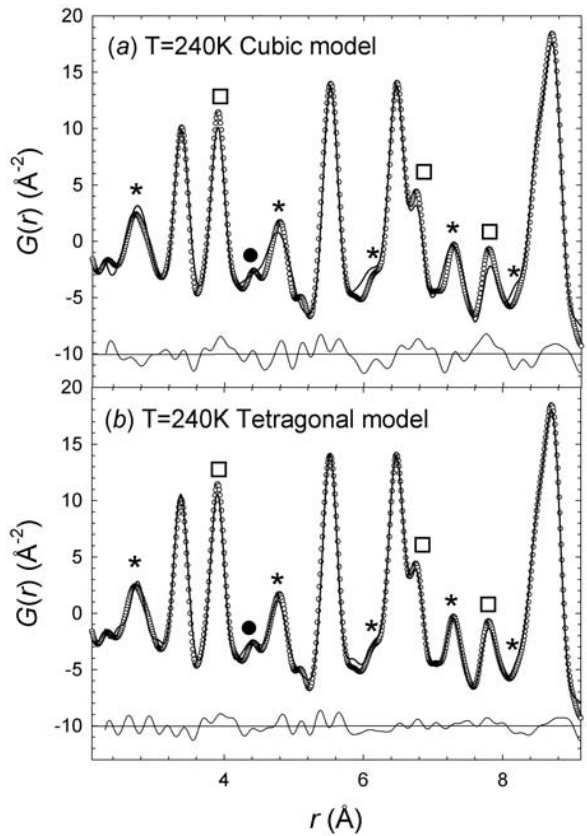

FIG. 8. Short range observed (dots) and calculated (continuous lines) PDF for ETO at 240K. (*a*) and (*b*) are the fits obtained using the cubic *average* model and low temperature tetragonal model, respectively. The symbols in the panel label the PDF peaks belonging to Eu-O (stars), Eu-Eu (empty squares), Ti-O (full circles) interatomic distances.